\documentstyle[12pt,aps]{revtex}

\newcommand{\dd}{{\rm d}}
\begin{document}
\tighten
\draft

\title{\flushleft Comment on `Self-dressing and radiation reaction in
       classical electrodynamics'\footnote{This comment is written by
V Hnizdo in his private capacity. No official support or endorsement by 
the Centers for Disease Control and Prevention
is intended or should be inferred.}}

\author{\flushleft V Hnizdo}

\address{\flushleft National Institute for Occupational Safety
and Health, 1095 Willowdale Road, Morgantown,\\ WV 26505, USA}                                      
\maketitle

\begin{abstract}
\noindent
Using the canonical formalism, Compagno and Persico  
(2002 {\it J.\ Phys.\ A: Math.\ Gen.} {\bf 35} 3629--45) have calculated the  
`radiation-reaction' force on a uniform spherical charge
moving rigidly, slowly and slightly from its position at the time when 
the transverse electric field is assumed to vanish.
This force is shown to result in the same time-averaged self-force
as that which has been obtained by different means for the test 
charge of a Bohr--Rosenfeld field-measurement procedure 
and which Compagno and Persico claimed to be incorrect.
\end{abstract}

In a recent paper \cite{CP}, Compagno and Persico (CP) have calculated,
by solving the coupled charge--field Hamilton equations of motion, 
the `radiation-reaction' force on a spherically
symmetric charge that moves rigidly, slowly and only a little from its position  
at the time when the transverse electric field is assumed to vanish.
For a charge $q$ that is uniformly distributed within a sphere of radius $a$, 
they obtain a `radiation-reaction' force 
\begin{equation}
F_{\rm RR}(t)= -\frac{2q^2}{a^3}\int_0^t\dd t'\,\dot{Q}(t')
\left[1-\frac{3(t-t')}{2a}+\frac{(t-t')^3}{4a^3}\right]\Theta[2a-(t-t')]
\label{FRR}
\end{equation}
where $\dot{Q}(t)$ is the time derivative      
of a one-dimensional trajectory of the charge and $\Theta(x)$ 
is the Heaviside step function;
here and henceforth, we use units such that the speed of light $c=1$ and put
$t=0$ for the time at which the transverse electric field  
vanishes. 

CP remark that result (\ref{FRR}) is relevant to the issues raised
in recent papers \cite{CP1,VH1,CP2,VH2} in connection with
the Bohr--Rosenfeld (BR) analysis of the measurability of the electromagnetic 
field \cite{BR}, as it should apply to a BR  measurement 
procedure with only minor modifications. 
In this comment, we show that the force (\ref{FRR}), 
which we prefer to call the electromagnetic self-force, results  
in the same time-averaged self-force as that which has been obtained 
by different means 
in \cite{VH1,VH2} for the test charge of a BR measurement procedure 
and which CP have rejected in \cite{CP2} as incorrectly calculated.

A condition on the one-dimensional trajectory $Q(t)$ of the test charge
in a BR measurement procedure occupying a time interval $(0,T)$ is that 
$Q(t)=0$ for $t\le 0$. While this means that the transverse electric field of 
the test charge vanishes at $t=0$, the initial condition
$\dot{Q}(t)|_{t=0}=0$ that is implied would result
according to (\ref{FRR}) in $Q(t)=0$ also for $t>0$ if there were 
no other force acting on the test charge in addition 
to the self-force. It is presumably for this reason that when CP
touch on the applicability of (\ref{FRR}) to the test
charge of the BR measurement procedure, they invoke the
`neutralization' of the test charge at $t=0$ by the stationary neutralizing
charge employed in the procedure. Besides the question whether such a 
neutralization alone indeed guarantees a vanishing transverse electric 
field at $t=0$, the fact remains that then 
there is at least one other force acting on the test charge, namely the 
electrostatic force of attraction to the neutralizing charge. It thus 
appears inescapable that any meaningful use of formula (\ref{FRR}) in
an analysis of the BR field-measurement procedure requires the
presence of external forces. We shall first write (\ref{FRR}) in a
different form before returning to this point.  

Using integration by parts, we write (\ref{FRR}) as
\begin{equation}
F_{\rm RR}(t)=Q(t')u(t-t')\Big|_{t'=0}^t-\int_0^t\dd t'\,Q(t')
\frac{\dd u(t-t')}{\dd t'}
\label{pparts}
\end{equation}
where 
\begin{equation}
u(t-t')=-\frac{2q^2}{a^3}\left[1-\frac{3(t-t')}{2a}+\frac{(t-t')^3}{4a^3}      
\right]\Theta[2a-(t-t')].
\label{u}
\end{equation}
Now, $u(t-t')|_{t'=t}=-2q^2/a^3$ and 
\begin{equation}
\frac{\dd u(t-t')}{\dd t'}
=-\frac{3q^2}{2a^4}\left[2-\frac{(t-t')^2}{a^2}\right]
\Theta[2a-(t-t')]. 
\label{u'}
\end{equation}
Using this and an initial condition $Q(t')|_{t'=0}=0$, we obtain 
the self-force (\ref{pparts}) for $t<T$ as
\begin{equation}
F_{\rm RR}(t)=-\frac{2q^2}{a^3}Q(t)+\frac{3q^2}{2a^4}\int_0^T\dd t'\,Q(t')
\left[2-\frac{(t-t')^2}{a^2}\right]\Theta(t-t')\Theta[2a-(t-t')] 
\label{F}
\end{equation}
where the factor $\Theta(t-t')$ is introduced in the integrand
in order to be able to fix the integration range as $(0,T)$.
Apart from differences in notation, this expression is the same 
as that for the self-force $F_x(t_2)$ 
given in \cite{VH2} by equations (26) and (28), with a factor of 2
instead of 3 in the delta-function term of equation (28) so that
the electrostatic force due to a BR neutralizing charge is not
included.  

The self-force $F_x(t_2)$ was obtained in \cite{VH2} by using the 
electromagnetic self-field of a uniform spherical charge whose trajectory 
was {\it prescribed} to be a given trajectory $Q(t)$ satisfying
the conditions imposed on it by the BR field-measurement procedure (but not
necessarily the stipulation that it is to have a step-like character with
respect to the measurement period $(0,T)$).
This fact invalidates the caution of CP that the force which they
obtained `is unambiguously the radiation-reaction force only in the
absence of other forces'. A prescribed trajectory
can eventuate to a given degree of accuracy only when there is a
suitable external force acting on the charge in addition to the self-force.  
In the BR field-measurement procedure, the test charge is acted on by an
external force that is the resultant of forces originating from several 
sources: the momentum-measurement system, the neutralizing charge,
the spring that compensates the time-averaged effects of the self-force and 
the neutralizing charge, and, of course, the measured external field itself; 
the approximate attainment of the prescribed step-like trajectory 
is there facilitated  by the fact that the test charge is allowed to have 
an arbitrarily great mass.

The self-force averaged over a time $T$,
\begin{equation}
\bar{F}_{\rm RR}=\frac{1}{T}\int_0^T\dd t\,F_{\rm RR}(t)
\label{Fbar}
\end{equation}
can now be written using (\ref{F}) as 
\begin{equation}
\bar{F}_{\rm RR}=-\frac{2q^2}{a^3T}\int_0^T\dd t\, Q(t)
+\frac{q^2}{T}\int_0^T\dd t'\, Q(t')\int_0^T\dd t\, v(t-t')
\label{Fbarv}
\end{equation}
where
\begin{equation}
v(t-t')=\frac{3}{2a^4}\left[2-\frac{(t-t')^2}{a^2}\right]
\Theta(t-t')\Theta[2a-(t-t')].
\label{v}
\end{equation}
The integration with respect to $t$ in the second term of (\ref{Fbarv}) 
is straightforward, yielding 
\begin{equation}
\int_0^T\dd t\, v(t-t')=\frac{2}{a^3}+f(t')
\label{intv}
\end{equation}
where
\begin{equation}
f(t')=-\frac{1}{2a^3}(2-\chi)(2-2\chi-\chi^2)\Theta(2-\chi)
\;\;\;\;\;\;\;\;\;\chi=\frac{T-t'}{a}.
\label{f}
\end{equation} 
Using (\ref{intv}) in (\ref{Fbarv}), the time-averaged self-force
is obtained finally as
\begin{equation}
\bar{F}_{\rm RR}=\frac{q^2}{T}\int_0^T\dd t'\,Q(t')f(t').
\label{Fb}
\end{equation}
Expressions (\ref{f}) and (\ref{Fb}) are, apart from  differences 
in notation and the absence in (\ref{f}) of the electrostatic term $-1/a^3$ 
due to the neutralizing charge, the same as 
those for the time-averaged self-force on the test charge of a BR-like
measurement procedure given in equations (3) 
and (9) of \cite{VH1} and equations (11) and (22) of \cite{VH2}; 
these expressions have been claimed by CP to be incorrect \cite{CP2}.

It has been shown in \cite{VH2} that  
the rejection \cite{CP2} of CP of the criticism \cite{VH1} 
of their re-analysis  \cite{CP1} of the BR field-measurement procedure
is based on erroneous calculations.  
It has to be concluded  that this rejection is invalidated also by
the results of the latest work of CP themselves.


\begin{references}

\bibitem{CP} Compagno G and Persico F 2002 Self-dressing and radiation reaction 
             in classical
             electrodynamics {\it J.\ Phys.\ A: Math.\ Gen.} {\bf 35} 
             3629--45
\bibitem{CP1} Compagno G and Persico F 1998 Limits on the measurability
of the local quantum electromagnetic-field amplitude {\it Phys. Rev.} A 
{\bf 57} 1595--1603
\bibitem{VH1} Hnizdo V 1999 Comment on Limits of the measurability of the 
local quantum electromagnetic-field amplitude {\it Phys. Rev.} A {\bf 60} 
4191--95; quant-ph/0210074
\bibitem{CP2} Compagno G and  Persico F 1999 Reply to Comment on Limits 
of the measurability of the local quantum electromagnetic-field amplitude 
{\it Phys. Rev.} A {\bf 60} 4196--97 
\bibitem{VH2}Hnizdo V 2000 The electromagnetic self-force on a charged
    spherical body slowly undergoing a small, temporary displacement from
    a position of rest {\it J.\ Phys.\ A: Math.\ Gen.} {\bf 33} 4095--103;
    math-ph/0005014
\bibitem{BR} Bohr N and Rosenfeld L 1933 Zur Frage der Messbarkeit der
elektromagnetischen Feldgr{\"o}ssen  {\it Mat. Fys. Medd. K. Dan. 
Vidensk. Selsk.} {\bf 12} no 8 (Engl. translation 1983 On the question of 
the measurability of electromagnetic field quantities 
{\it Quantum Theory and Measurement} ed J A  Wheeler and
W H Zurek (Princeton NJ: Princeton University Press) pp 479--522)

\end{references}
\end{document}